\documentclass[useAMS,usenatbib]{mn2e}

\usepackage{graphicx}
\usepackage{subfig}
\usepackage{epstopdf}
\usepackage{color}

\title[Microlensing of the Electron Scattering Region]{Gravitational Microlensing as a probe of the
Electron Scattering Region in Q2237+0305}
\author[D. J. Kedziora, H. Garsden \& G. F. Lewis]{D. J. Kedziora\thanks{E-mail:
d.kedziora@physics.usyd.edu.au (DJK);  hgarsden@physics.usyd.edu.au (HG); gfl@physics.usyd.edu.au (GFL)}, H. Garsden\footnotemark[1] and G. F. Lewis\footnotemark[1]\thanks{Research undertaken as part of the Commonwealth Cosmology 
Initiative (CCI: www.thecci.org), an international collaboration 
supported by the Australian Research Council.}\\
Sydney Institute for Astronomy, School of Physics, A28, University of Sydney, NSW, 2006, Australia}
\begin{document}

\date{Draft: September 2010}

\pagerange{\pageref{firstpage}--\pageref{lastpage}} \pubyear{2010}

\maketitle

\label{firstpage}

\begin{abstract}
Recent observations have provided strong evidence for the presence of an Electron Scattering Region (ESR) within the central regions of AGNs. This is responsible for reprocessing emission from the accretion disk into polarised radiation. The geometry of this scattering region is, however, poorly constrained. In this paper, we consider the influence of gravitational microlensing on polarised emission from the ESR in the quadruply imaged quasar, Q2237+0305, demonstrating how correlated features in the resultant light curve variations can determine both the size and orientation of the scattering region. This signal is due to differential magnification between perpendicularly polarised views of the ESR, and is clearest for a small ESR width and a large ESR radius. Cross- and auto-correlation measures appear to be independent of lens image shear and convergence parameters, making it ideal to investigate ESR features. As with many microlensing experiments, the time-scale for variability, being of order decades to centuries, is impractically long. However, with a polarization filter oriented appropriately with respect to the path that the quasar takes across the caustic structure, the ESR diameter and radius can be estimated from the auto- and cross-correlation of polarized light curves on much shorter time-scales.
\end{abstract}


\section{Introduction}\label{Section:Introduction}
While quasars are amongst the most luminous sources in the Universe, being at cosmological distances ensures that their small angular size renders them unresolved to modern telescopes. While direct imaging of the central engines of quasars is not possible, more novel techniques are able to reveal the scales of structure of several prominent features [e.g. reverberation mapping can reveal the size of the enveloping Broad-Line Region; see \citet{1998AdSpR..21...57P}]. Gravitational microlensing is one such method, occurring when the light from a distant source is magnified by the presence of stellar masses crossing the line of sight. Within the Galaxy, where single or binary stars pass in front of more distant sources, microlensing has been used to great effect in studying the surface properties of distant stars, such as limb darkening \citep[e.g.][]{Rhie:99}.

Microlensing also occurs in a cosmological context, where the light from a distant quasar has been multiply imaged by a foreground galaxy. As with the Galactic case, stars within the lensing galaxy which cross the line of sight to these images can induce substantial magnification effects. Unlike the Galactic case, the number of stars influencing a particular image can be many thousands, and hence the resultant pattern of magnification can be extremely complex \citep[see][for examples]{1986A&A...166...36K,Wambsganss:1990b}. Regions within the quasar will respond differently to the magnifying stars, dependent upon their size and location, and hence the light curve variations as the stars cross the line of sight encode the
underlying quasar structure. Understanding and, more importantly, deconvolving this encoding has been the focus of a substantial amount of study in recent years, and has included the development of computational approaches to account for the gravitational
lensing by a large population of sources \citep[see][for a recent review]{2006glsw.book.....S}. Microlensing has been used to study structure on a range of scales within quasars, both limiting the size of the central accretion disk \citep{Yonehara:2001,Bate:2008,Pooley:2010} and including the properties of the ``big blue bump'' \citep{Wambsganss:90a}. On larger scales, microlensing of the broad line region has been the subject of theoretical  \citep{1988ApJ...335..593N,1990A&A...237...42S,Abajas:2002,Lewis:2004} and observational \citep{Richards:2004,Wayth:2005} investigations, with differential microlensing between the inner disk and extensive line emitting region providing further clues to quasar structure \citep[e.g.][]{Keeton:2006}.

In this paper, we will consider the influence of microlensing on the prominent electron scattering region (ESR) responsible for the reprocessing of quasar accretion disk emission into polarised light \citep{Kishimoto:08a,Kishimoto:08b}. Although the use of polarimetry in probing the Broad Absorption region through microlensing has been suggested \citep{2000PASP..112..320B,Hales:07},
here we consider the influence of microlensing on the scattering region in general. Evidence for the microlensing of an electron scattering region in the gravitationally lensed quasar H1413+1143 has been provided by \citet{Chae:2001}, where it was found that one of the images had a rotated polarisation position angle and a somewhat higher degree of polarisation than the other images; this was attributed to microlensing of an ESR.

The focus of this paper is to examine the influence of gravitational microlensing on fiducial models of the ESR, using numerically generated maps of the complex magnification patterns found in quasar microlensing. The structure of this paper is as follows;
we detail our model of an electron scattering region in Section \ref{Section:Background}, as well as describe the mathematical
formalism behind gravitational microlensing and the method by which we apply this to our model. Section \ref{Section:Results} presents the results of cumulative and time-series statistical analyses with the aim of discovering methods of identifying the size and orientation of an ESR. The conclusions are presented in Section \ref{Section:Conclusion}.

\section{Background and Approach}
\label{Section:Background}

\subsection{Source Model}
\label{Subsection:Model}

\begin{figure}
\centering
\includegraphics[trim = 0mm 0mm 0mm 0mm, clip, width=0.52\textwidth]{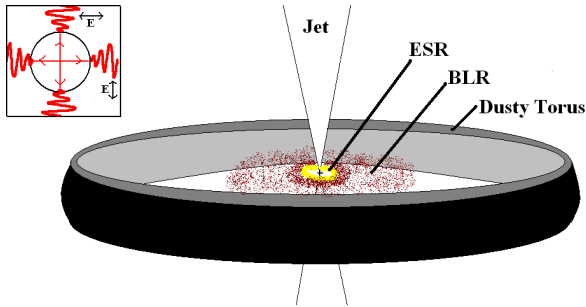}
\caption{Schematic model of an AGN viewed from the side, with the jet-emitting core and accretion disk surrounded by an electron scattering region. Beyond this is a broad line emission region that may overlap the ESR and is itself surrounded by a dusty torus. In the top left corner is a top-view schematic that depicts how unpolarised light from the central source becomes polarised after scattering by the ESR; the arrows represent the direction of $E$ polarisation.}
\label{Fig:QuasarSchematic}
\end{figure}

Quasars represent the most powerful of active galactic nuclei, with emission arising from very energetic compact regions thought to be supermassive black holes powered by accretion disks. Although there are several models for the specific aspects of quasar structure \citep[e.g.][]{Elvis:00,Schild:03}, the consensus is that a broad line emitting region (BLR) lies outside the accretion disk and a dusty torus exists beyond that, both of which fuel the central accretion disk. The dusty torus is typically considered to be on the scale of $0.1$ kiloparsecs, but sizes for both the BLR and torus are not well constrained. Recent publications about the dusty torus have even argued for a minimum radius on the order of parsecs \citep[e.g.][]{Agol:09}.

It has been suggested that an additional electron scattering region (ESR) is required to explain polarized emission from quasars \citep{Taniguchi:99,Kishimoto:08a} as polarimetry appears to cut through the scattered emission of the dusty torus, implying that scattering from the interior is able to produce spectra of the accretion disk without contamination from dust emission. However, the geometry of the ESR and its position relative to the BLR is still a subject of debate \citep{Kishimoto:08b}; a combined model possessing all of the key features within a quasar is displayed in Figure \ref{Fig:QuasarSchematic}.

The quasar's accretion disk is represented by a 2D Gaussian of radius
\begin{equation}
\label{Eq:HalfWidthRadius}
r_{1/2}=4 \times 10^{14} \left(\lambda / \mu m \right)^{1.5}\ m,
\end{equation}
where $r_{1/2}$ is the half-light radius of the accretion disk and $\lambda$ is the emission wavelength in the quasar's rest frame \citep{Agol:09}. For the purpose of this investigation, $\lambda=0.7$ $\mu$m (i.e. infrared) is used to comply with the ESR's scattering wavelengths of interest \citep{Kishimoto:08b}. The accretion disk's presence is primarily to place a lower bound on the ESR radius.

For the purpose of this study, we ignore the BLR as, although polarisation levels from this region are potentially quite high \citep{Antonucci:99}, it should be possible to discount its effects due to the presence of emission line spectral signatures \citep{Kishimoto:08a}. The dusty torus is also neglected in our model due to its immense size, meaning that it will be largely unaffected by microlensing, producing intensity variations of less than $1\%$ \citep[e.g.][]{Refsdal:91}. In contrast, the ESR is considered to be contained within distances below the minimum radius of the dusty torus and is thus far more susceptible to small-scale magnification variations that affect its intensity.

\begin{figure}
\centering
\includegraphics[trim = 0mm 0mm 0mm 0mm, clip, width=0.48\textwidth]{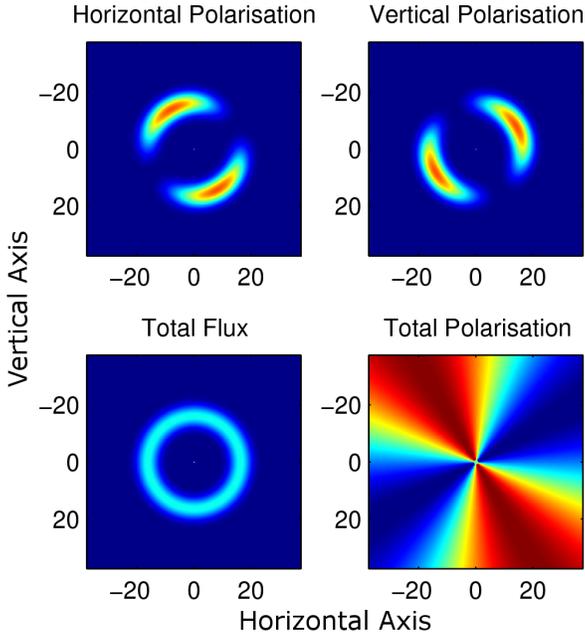}
\caption{An example of a simple ESR model with minimum radius of $0.64$ pc and a width of $0.64$ pc. The top left image is the ESR viewed through a HP filter rotated anti-clockwise by $\pi/6$, i.e. $\theta_f = \pi/6$ in Equation \ref{Eq:ModelEquation}. If the polarisation filter was not rotated, i.e. if $\theta_f = 0$, the wings would lie horizontally. The top right image is the ESR viewed through a VP filter rotated anti-clockwise by $\pi/6$. Without a rotated filter, the wings would lie vertically. Red and blue in these two panels represent high and low intensities, respectively. The bottom left image is the total flux of the ESR. The bottom right image is the total polarisation of the ESR as defined in Equation \ref{Eq:TotalPol}. Red and blue in this panel represent high and low levels of $\mathrm{Pol}_{\mathrm{Tot}}$, respectively. Axis units are in Einstein radii for Q2237+0305 (see Section~\ref{Subsection:Mathematics}).}
\label{Fig:ModelExample}
\end{figure}

As a fiducial model, we assume the case where the quasar jet is aimed at the observer and the ESR is seen as a rotationally symmetric, radially Gaussian annulus; while simple, this model captures the key physical features of the ESR, although we intend to generalize this model in future contributions. With this assumption, the source is an extreme-case Type 1 object, where the nuclear region interior to the dusty torus is visible. The polarisation $E$ vector of scattered emission is tangent to the circumference of the annulus (as in the insert, Figure \ref{Fig:QuasarSchematic}) and, in the case of a face-on disk, any circularly symmetric ESR will result in zero net polarisation. However, due to the effects of differential magnification, parts of the polarised emission will be preferentially magnified, resulting in a net polarisation. A polarising filter may then be used to view the annulus such that either the horizontal or vertical component of emission is detected; these will be called the ``horizontally polarised'' (HP) and ``vertically polarised'' (VP) filters. The filter may then be rotated, admitting light
polarised at an angle from the horizontal or vertical. These points are summarised in the following equation, describing the visible ESR intensity profile $I$ as a function of polarisation filter and orientation, annulus dimensions, annulus intensity and location $\theta$ over the source plane:
\begin{equation}
\label{Eq:ModelEquation}
I=I_0\ \sin^2 ( \theta + \theta_{a} - \theta_{f} )\ \exp\left(-\frac{(r-(r_E+w_E/2))^2}{2(w_E/4)^2}\right),
\end{equation}
where $r$ is the radial distance from the quasar centre, $r_E$ is the minimum radius of the ESR, $w_E$ is the width of the ESR, $I_0$ is the flux normalisation, $\theta$ is the source plane location, $\theta_{a}$ indicates use of a horizontal or vertical filter, and $\theta_{f}$ is the rotation of the filter. The third factor in the equation implements the Gaussian cross-section of the annulus, which peaks at the average radius of the ESR and emits $95\%$ of its total photon flux from within its ``width''. The second factor implements the effect of a polarising filter. A horizontal or vertical filter is specified by $\theta_\alpha = 0$ or $\pi/2$ respectively, and rotation of the filter is specified with a non-zero value of $\theta_f$. In all cases the visible source profile is a pair of mirror-image ``wings'', as depicted in Figure \ref{Fig:ModelExample}. The top left image is the ESR viewed through the HP rotated filter; the wings would be horizontal except for the fact that the filter is rotated by angle $\theta_f = \pi/6$, leading to a corresponding profile rotation. The top right image uses a rotated VP filter; the wings would be vertical except for the fact that the filter is at $\theta_f = \pi/6$. In these top two panels, red and blue indicate high and low intensities, respectively. The orientation of the profile is therefore specified by one of HP or VP plus the angle $\theta_f$. The bottom left image is the total flux profile for the annulus, and the bottom right image is the total polarisation, defined as
\begin{equation}
\label{Eq:TotalPol}
\mathrm{Pol}_{\mathrm{Tot}} = \frac{I_{\mathrm{HP}} - I_{\mathrm{VP}}}{I_{\mathrm{HP}} + I_{\mathrm{VP}}}
\end{equation}
at $\theta_f = \pi/6$, where $I_{\mathrm{HP}}$ and $I_{\mathrm{VP}}$ are the intensities of the horizontally and vertically polarised components, respectively. In this panel, red and blue indicate high and low values of $\mathrm{Pol}_{\mathrm{Tot}}$, respectively.

\subsection{Gravitational Lensing}
\label{Subsection:Mathematics}

The formalism of microlensing is well described in the literature \citep[see][]{Schneider:92}; however, we reproduce the key aspects here. Strong gravitational lensing occurs when the light from a distant source is deflected by a massive object, allowing several light paths to connect it to an observer and resulting in multiple images of the distant source. For galaxy-mass lenses, the typical separation of these images are of order an arcsecond, although individual stars within the lensing galaxy can also induce additional deflections as they pass through the line-of-sight to a source. While these deflections, of order a micro-arcsecond, are too small to produce additional resolvable images, they can produce substantial brightness variations; this is the observable consequence of gravitational microlensing. 
  
For this study, the key characteristic length that arises in gravitational lensing is the Einstein Radius; this describes the radius of a ring seen around a point-mass lens when there is perfect alignment between the source, lens and observer. Projected onto the source plane, this is given by
\begin{equation}
\label{Eq:EinsteinRadius}
\zeta_0=\sqrt{\frac{4 G M}{c^2} \frac{D_{os} D_{ls}}{D_{ol}}},
\end{equation}
where $M$ is the mass of the lens and $D_{xy}$ refers to the angular diameter distance between $x$ and $y$; the subscripts $s, l,$ and $o$ representing source, lens, and observer respectively. 

While gravitational lenses are three-dimensional mass distributions, the cosmological distances involved mean that in practice
we can treat them with the thin lens approximation, where the mass distribution is projected onto a lens plane, giving a surface mass density $\Sigma(\xi)$ at point $\xi$ in the plane. This is normalised by a critical surface mass density of
\begin{equation}
\label{Eq:CriticalSurfaceMassDensity}
\Sigma_{cr}=\frac{c^2}{4 \pi G}\frac{D_{os}}{D_{ol} D_{ls}},
\end{equation}
such that the normalised surface mass density is given by $\sigma = \Sigma / \Sigma_{cr}$, also referred to as convergence ($\sigma$). The mapping of intensity between the source and image planes can be expressed with a Jacobian matrix of the form   
\begin{equation}
\label{Eq:JacobianMatrix}
A=\left (\begin{array}{cc} 1-\sigma-\gamma_1 & -\gamma_2 \\ -\gamma_2 & 1-\sigma+\gamma_1 \end{array} \right ),
\end{equation}
where  $\gamma=\sqrt{\gamma_1^2+\gamma_2^2}$ is called the shear and incorporates the long-range influence of the lensing mass distribution.

\begin{figure}
\centering
\subfloat{\label{Fig:MagMap1}\includegraphics[trim = 0mm 0mm 0mm 0mm, clip, width=0.23\textwidth]{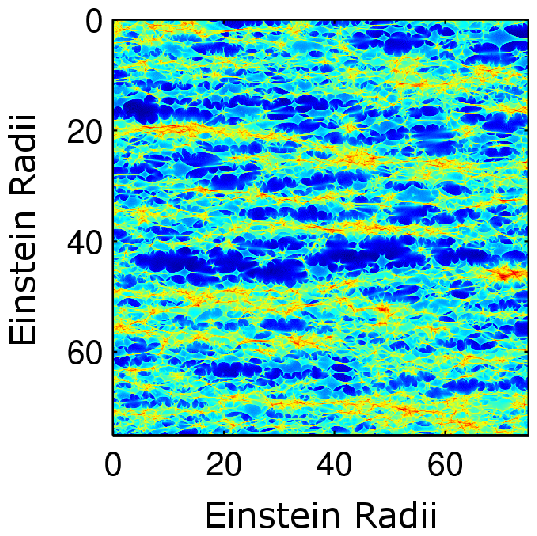}}
\subfloat{\label{Fig:MagMap2}\includegraphics[trim = 0mm 0mm 0mm 0mm, clip, width=0.23\textwidth]{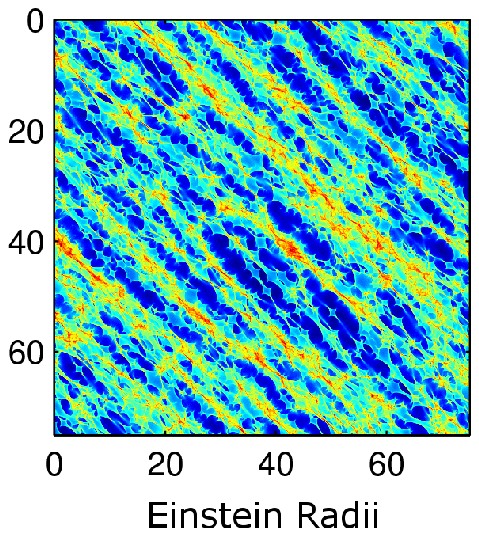}}
\caption{Two different examples of magnification maps for image B with $\sigma=0.38$ and $\gamma=0.43$. The second one is rotated by $\phi = -\pi/4$. Spatial units are in Einstein Radii for a solar mass star.}
\label{Fig:MagMapExample}
\end{figure}

The theoretical magnification of an image is then given as
\begin{equation}
\label{Eq:TheoreticalMagnification}
\mu_{th}=(\det A)^{-1}=\frac{1}{(1-\sigma)^2-\gamma^2}.
\end{equation}
Note that formally infinite magnifications are possible when the denominator of Equation \ref{Eq:TheoreticalMagnification} goes to zero. This defines a network of critical curves in the image plane that map to caustics in the source plane; patterns of light and dark seen in Figure \ref{Fig:MagMapExample} \citep[see][for examples]{1990A&A...236..311W}, described in full later in the text.

For convenience it is common to set $\gamma_2$ to zero, thus choosing an orientation where the dominant caustic structures are considered to be horizontal \citep[as an example, see Figure 2 in][]{Wambsganss:1990b}. However, rotation of the caustic network is possible by noting
\begin{equation}
\label{Eq:MagmapRotation}
\tan(2 \phi)=\frac{\gamma_2}{\gamma_1},
\end{equation}
where $\phi$ represents the angle between the dominant caustic structures and the horizontal axis (Figure 
\ref{Fig:MagMapExample}). We will henceforth refer to these structures as caustic ``bands''.

\begin{table}
\begin{center}
\medbreak
\begin{tabular}{cccc}
\hline Image & $\sigma$ & $\gamma$ & $\mu_{th}$ \\
\hline A & $0.41$ & $0.47$ & $(+)7.86$\\
 B & $0.38$ & $0.43$ & $(+)5.01$\\
 C & $0.65$ & $0.68$ & $(-)2.94$\\
 D & $0.59$ & $0.56$ & $(-)6.87$\\
\hline
\end{tabular}
\caption{Convergence, shear and mean theoretical magnification for the four images seen in the gravitational lens Q2237+0305  \citep[taken from the best-fit model of][]{1992AJ....104..959R}.\label{Tab:LensParameters} The sign of the magnification denotes parity.}
\end{center}
\end{table}

In practice, each image of a multiply-imaged lensed quasar is modelled with a convergence and shear value describing the image magnification within the overall model of the system. As the lens and source move relative to each other over time, microlensing by the stellar objects within the lens will deviate the magnification above and below this value. For the purposes of this study, we will focus upon the gravitational lens Q2237+0305 \citep{Huchra:85} due to its significant historical monitoring \citep{Wozniak:2000,Udalski:2006}, which has revealed ongoing microlensing. Q2237+0305 is an attractive laboratory for microlensing investigations because of the low redshift of the lensing galaxy, leading to high microlensing variability \citep{Kayser:89}. It is a quadruply imaged $z=1.69$ quasar seen through the bulge of a $z=0.04$ spiral galaxy; the Einstein Radius for a solar mass star in this system is $\sim 0.06$ pc. The adopted values of the shear and convergence for each image in Q2237+0305 are given in Table~\ref{Tab:LensParameters}.

Analysis of recent Q2237+0305 observations by the Spitzer Space Telescope used several models to constrain the minimal radius of the dusty torus \citep{Agol:09}. The largest value was calculated as $3.83$ pc for an interstellar medium model \citep{Draine:03}, so we treat this as the conservative upper bound on the ESR radius.

\subsection{Numerical Analysis of Gravitational Microlensing}
\label{Subsection:Method}
For this study, we used  backwards ray-tracing \citep{1986A&A...166...36K,Wambsganss:90,2010NewA...15..181G} to generate microlensing magnifications; here, rays are fired from an observer through a plane of lensing masses. These masses deflect the path of the rays, which are then traced into a distant source plane. The plane is represented as a pixel grid, with the number of rays that land in each pixel of the source plane corresponding to the magnification of a pixel-size source situated at that pixel. The result is a magnification map (Figure \ref{Fig:MagMapExample}) of the source plane, where bright areas indicate locations where the source will be highly magnified and dark areas indicate where they will be demagnified. Extended sources are represented as a flux profile pixel grid and convolved with the magnification map to produce a map for the microlensing of that source.

The analysis in this paper considers all microlensing masses to be $1$ Solar mass, with the Einstein Radius (ER) projected into the source plane of $0.06$ pc. Note that, as the Einstein Radius is proportional to $\sqrt{M}$, all distances related to this problem are scalable by this factor. While any realistic field of microlensing bodies is expected to possess a mass distribution, the overall statistical properties are proportional to the square root of the mean mass \citep[e.g.][]{Lewis:95a}.

All magnification maps are $150$ ER ($9$ pc) across, with the inner $75$ ER ($4.5$ pc) used so as to avoid edge effects. Considering that the ESR is expected to lie  $\sim 0.1$ pc away from the centre of the quasar \citep{Taniguchi:99}, we henceforth set the 
inner radius of the ESR to be $0.08$ pc ($1.33$ ER) and the annulus width to be $0.32$ pc ($5.33$ ER), unless otherwise stated. This model will be referred to as the ``base'' or fiducial ESR. To obtain a statistical sample, we generate $20$ maps for each of the four lensed images in Q2237+0305 (see Table \ref{Tab:LensParameters}), each map having the lensing masses at different randomly selected positions. Every map is also rotated by $\phi = 0$, $-\pi/12$, $-\pi/6$ and $-\pi/4$, as enabled by Equation \ref{Eq:MagmapRotation}.

\section{Results}
\label{Section:Results}

In the first subsection we present the results of magnification map analysis for different source profiles. The second
subsection studies the dynamical effects of microlensing by analysing light curves.

\subsection{Cumulative Statistics}
\label{Subsection:StaticStatistics}

\begin{figure}
\begin{center}
\subfloat[]{\label{Fig:LowHistogram}\includegraphics[trim = 0mm 0mm 0mm 0mm, clip, width=0.50\textwidth]{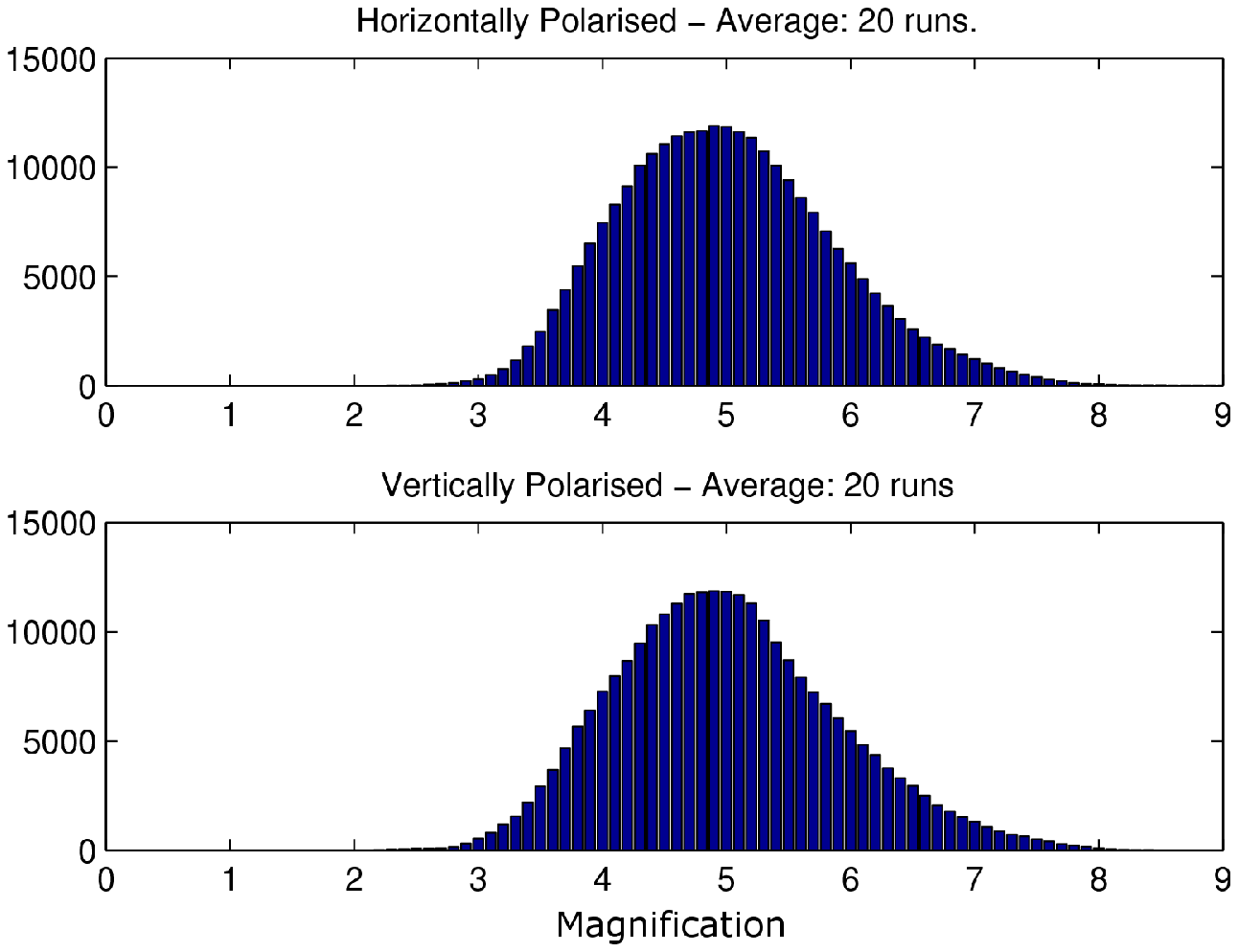}}\\
\subfloat[]{\label{Fig:HighHistogram}\includegraphics[trim = 0mm 0mm 0mm 0mm, clip, width=0.50\textwidth]{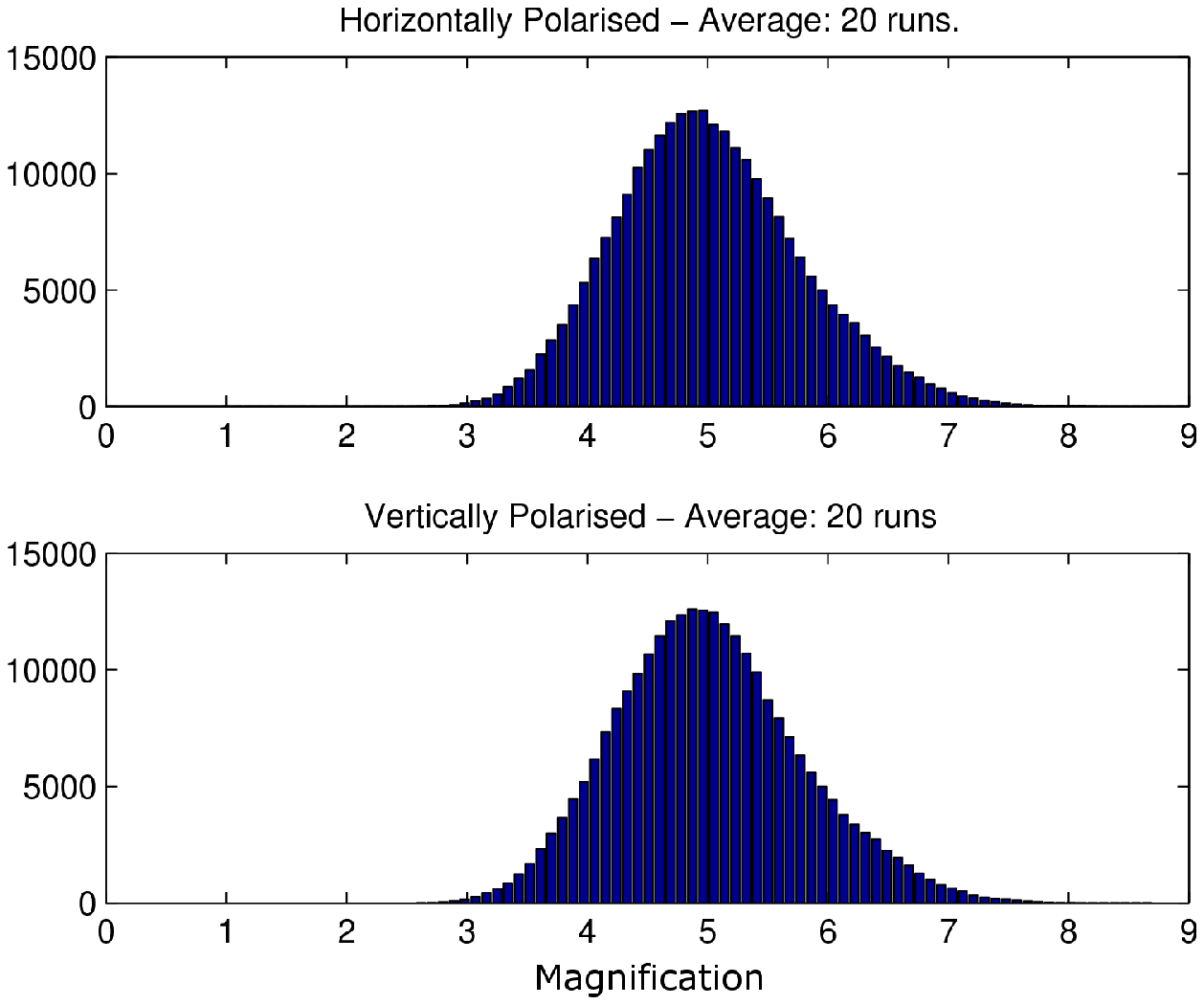}}\\
\subfloat[]{\label{Fig:DiffHistogram}\includegraphics[trim = 0mm 0mm 0mm 0mm, clip, width=0.50\textwidth]{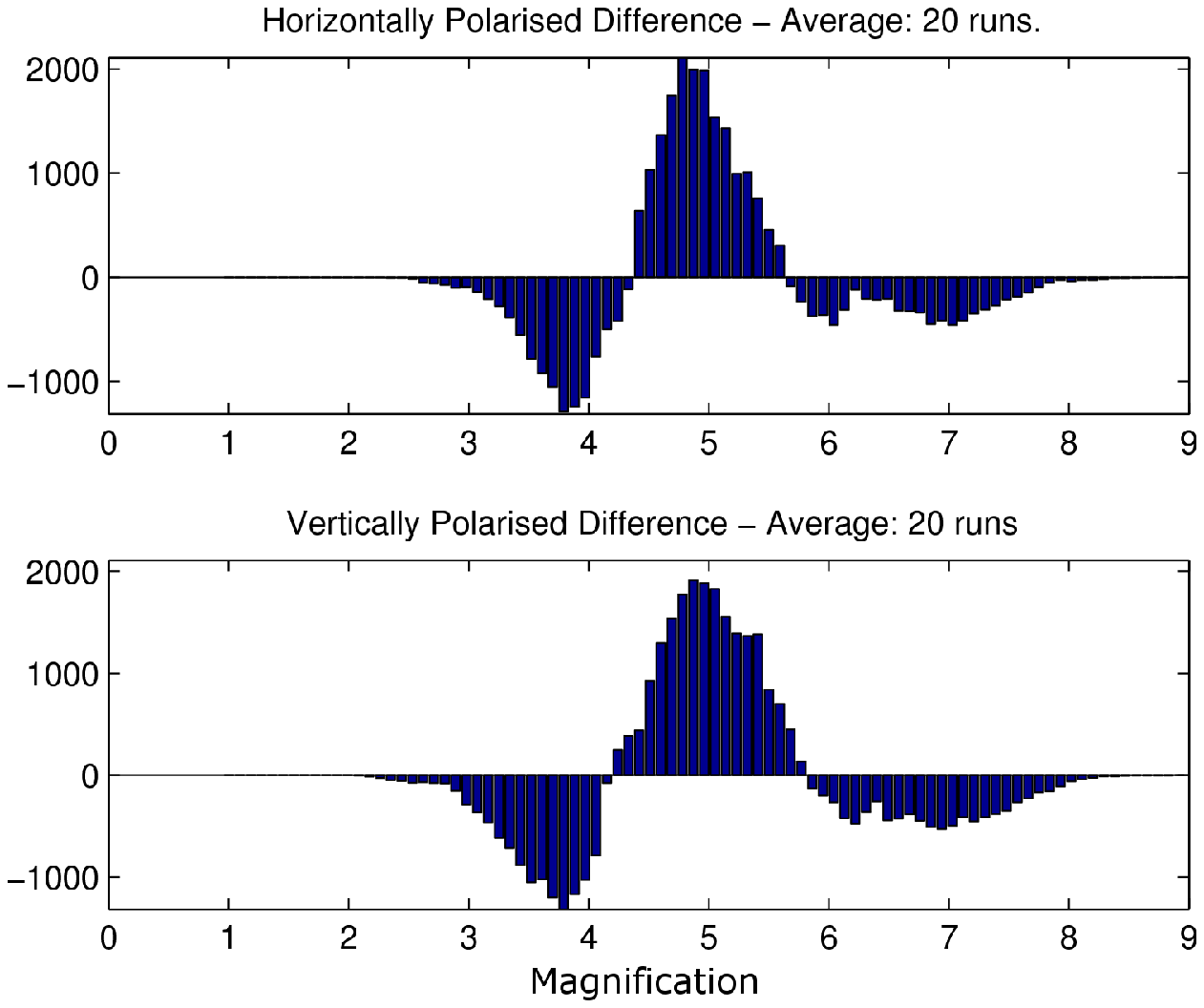}}
\caption{Magnification probability distributions (histograms) for Q2237+0305 B using the base ESR, unrotated maps. (a) represents the use of unrotated HP and VP filters. (b) represents a  rotation of $\theta_f = \pi/4$. (c) shows the additional number of counts that the histograms in (b) have over (a).}
\label{Fig:Histograms}
\end{center}
\end{figure}

Magnification probability distributions of magnification maps are shown in Figure \ref{Fig:Histograms} for the base ESR as seen in image B, using unrotated maps and as viewed through HP and VP filters at two filter orientations each; unrotated, and rotated by $\theta_f = \pi/4$. The distributions appear very approximately Gaussian in shape, centred on the expected theoretical magnification. In Figure~\ref{Fig:LowHistogram} the top and bottom panels use unrotated HP and VP filters, respectively. Figure \ref{Fig:HighHistogram} presents the same data but for $\theta_f = \pi/4$. By subtracting the former distribution from the latter, Figure \ref{Fig:DiffHistogram} shows that a relative angle of $\pi/4$ between polarisation filter and the band-like caustic structures, evident in Figure \ref{Fig:MagMapExample}, biases the distribution towards the mean with less extremes in magnification. However, subtracting the distribution in Figure \ref{Fig:HighHistogram} from a histogram for $\theta_f =\pi/2$ (not shown) generates a vertical inversion of Figure \ref{Fig:DiffHistogram}, with an increased proportion of high and low magnification with respect to the average. Further simulations confirm this trend for any additional rotation of $\pi/2$.

The explanation of this requires consideration of how the HP and VP wings overlap with the structure of the magnification map. If the relative orientation is such that either the HP or VP wings are parallel to the caustic bands, there is a high probability that either the wing pair crosses caustics and magnification is larger than average, or the wings are both off the caustic structure and magnification is below average. Presumably, a perpendicular orientation also leads to such extremes; if one wing lies across a caustic then the other is likely to cross the same band, and likewise if one wing is off the structure then the other is too. In contrast, a relative angle of $\pi/4$ does not double the magnification effect for a wing pair via simultaneous traversal of the same caustic, nor can the pair completely overlap a band as they do when parallel to the caustic structure. Hence the probability of magnification is more tightly centred around an average.

\begin{figure*}
\centering
\includegraphics[trim = 0mm 0mm 0mm 0mm, clip, width=0.58\textwidth]{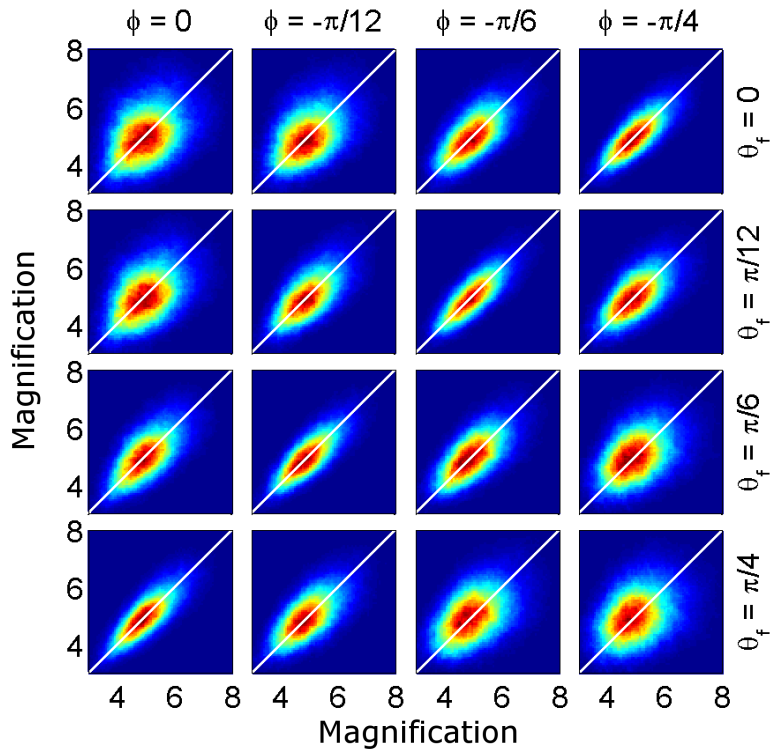}
\caption{Joint distributions for the magnifications of the base ESR seen through the VP filter (vertical axes) versus the magnifications seen through the HP filter (horizontal axes), all as seen in image B with varying map and filter rotations. Different columns correspond to various magnification map rotations and different rows correspond to various polarisation filter rotations. White lines denote equal magnification of both components.}
\label{Fig:RotationDistributionsB}
\end{figure*}

\begin{figure*}
\centering
\includegraphics[trim = 0mm 0mm 0mm 0mm, clip, width=0.58\textwidth]{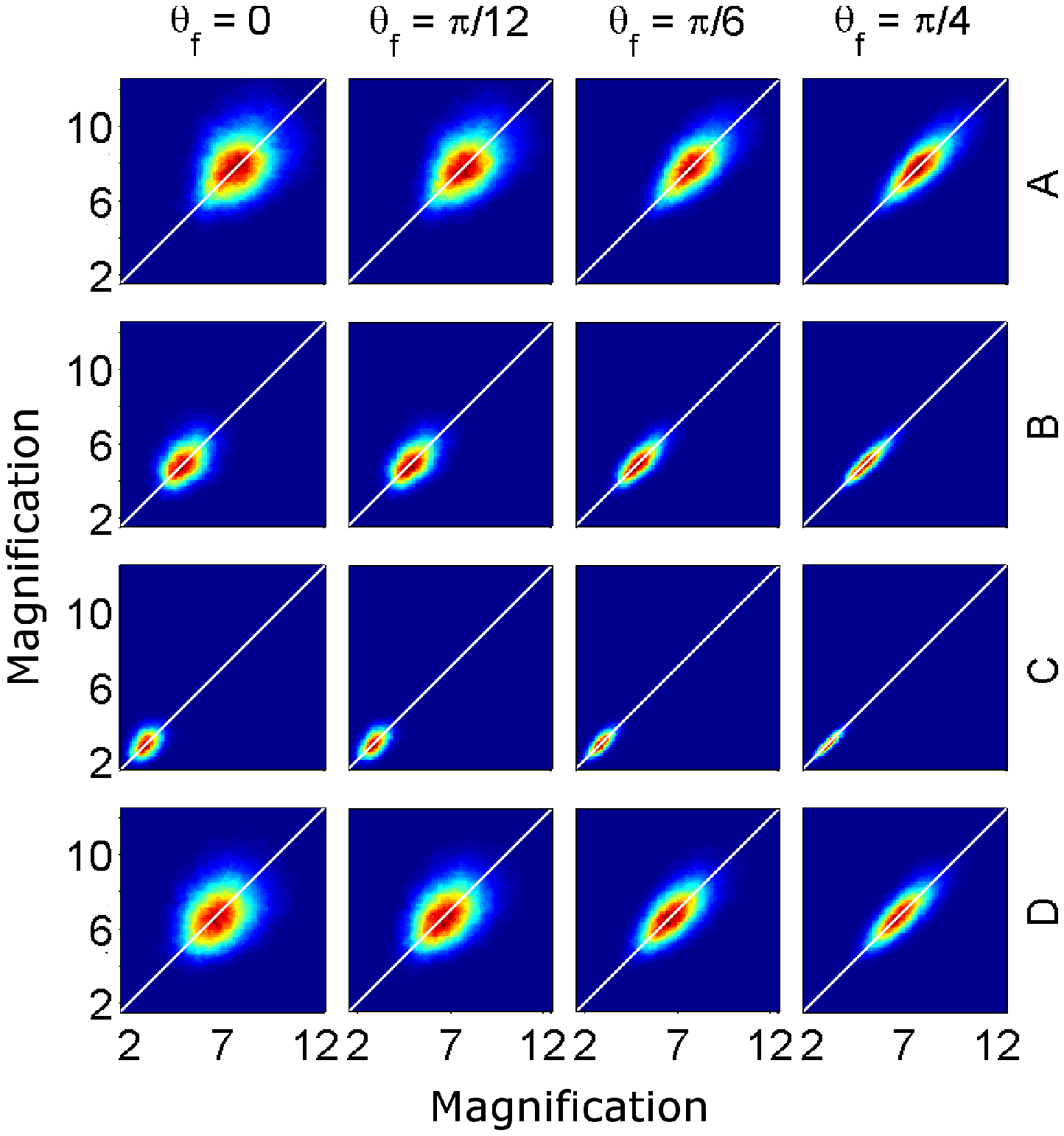}
\caption{Joint distributions for the magnifications of the base ESR viewed through the VP filter (vertical axes) versus the magnifications view through the HP filter (horizontal axes), with unrotated maps and for images A, B, C, D. Different columns correspond to various polarisation filter rotations and each row corresponds to a different image. White lines denote equal magnification of both components.}
\label{Fig:ABCDDistributionComparisons}
\end{figure*}

Figure \ref{Fig:RotationDistributionsB} extends this argument by displaying the joint probability distributions of the base ESR 
VP filter magnifications ($y$-axis) and HP filter magnifications ($x$-axis) for Q2237+0305, all as seen in image B at varying rotations of the polarisation filter and magnification maps. Joint distributions have been fruitfully used in other works \citep{Lewis:2004,Brewer:2005,Abajas:2007} and indicate the probability of HP and VP filtered magnifications occurring at the same source location, i.e. map point, and thus how well magnifications vary in ``lock-step''. They therefore indicate expectations of one  magnification based on the other, and can constrain models from observations. The probability value is indicated by colour, with blue being $0$ and red ``most likely''. All the panels represent the same measure, but for different orientations of the magnification map ($\phi$) and the various rotations applied simultaneously to both HP and VP filters ($\theta_f$).

All distributions appear symmetric across the lines of equal magnification. However, the top left and bottom right plots show widest distributions when the relative angle between filter and magnification map is a multiple of $\pi/2$. The spread of the distribution is a result of the extreme relative magnifications that occur when one wing pair of the ESR lands on a caustic and the other pair does not, be it parallel or perpendicular. In contrast to the $0$ and $\pi/2$ relative angle plots, the four on the diagonal from bottom left to top right represent a relative angle of $\pi/4$ between the VP wings and the shear axis. They show the greatest level of correlation between the magnification of horizontally and vertically polarised light as both wing pairs are symmetric across the axis perpendicular to the caustic bands; there is no position for the ESR with respect to the magnification map that allows significant overlap between the caustic structure and one wing pair but not the other.

Figure \ref{Fig:ABCDDistributionComparisons} replaces the variation in $\phi$ (here $\phi$ is $0$ everywhere) with the different lensed images. It shows that there is a tightening of the distribution around the line of equal magnification for relative angles of $\pi/4$ and this is a general effect for any of the four images. As expected, the distributions are also centred around the theoretical magnifications for each lens, with values presented in Table \ref{Tab:LensParameters}.

\begin{figure}
\centering
\includegraphics[trim = 0mm 0mm 0mm 0mm, clip, width=0.45\textwidth]{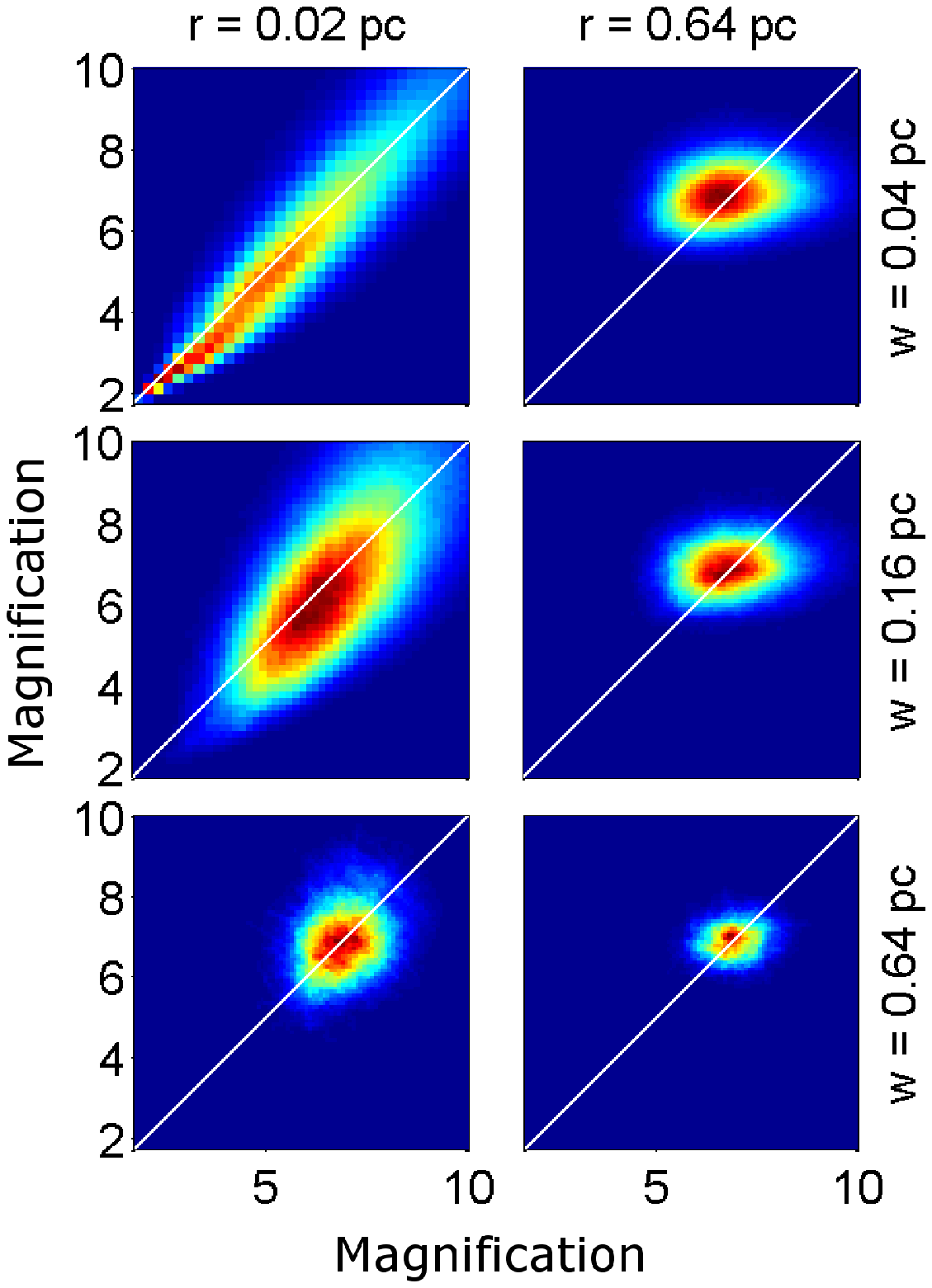}
\caption{Joint distributions for the magnifications using the VP filter (vertical axes) versus the magnifications using the HP filter (horizontal axes), all as seen in image D with unrotated maps and unrotated filters. The
size of the ESR model changes from panel to panel; the rows correspond to the changing inner radius of the ESR annulus and the columns to its width. White lines denote equal magnification of both components.}
\label{Fig:SizeComparisons}
\end{figure}

Thus far we have only commented on features that indicate the relative orientation between the shear axis of a lens and a polarisation filter. However, it is also clear that source size plays a role in the probability distribution of magnification. Figure \ref{Fig:SizeComparisons} shows distributions for image D with $\theta_f = \phi = 0$ and different ESR annulus sizes, indicated by $r$ and $w$. These sizes indicate that a small ESR on the low end of the assumed size range \citep{Taniguchi:99} has a long tail in its magnification distribution, suggesting large fluctuations in relative intensity. A heightened degree of correlation between observed horizontal and vertical polarisation intensities should be expected due to the small source size and the large amount of time the entire annulus is on or off a caustic relative to the amount of time a polarised component is. In contrast, a wide ESR annulus only ever has minor portions overlapping caustic bands, and thus variations in the intensity around the theoretical mean are much smaller.

For annuli of large inner radius, Figure \ref{Fig:SizeComparisons} shows that the distributions are tightened around a mean. The explanation for the change is similarly to do with the ESR area; the magnification map has a fixed characteristic length scale for caustics, as implied by Figure \ref{Fig:MagMapExample}, and thus a larger ESR will have an intensity that is more robust to perturbations in magnification as it moves relative to the lens.

We finally note that the last panel in Figure \ref{Fig:SizeComparisons} is an analogy to the dusty torus situation that we excluded from the model in Section \ref{Subsection:Model}. There is a lack of correlation between the VP and HP regions due to their size, and magnification variability is on the order of only $5\%$. Hence, if the intensity of observed polarisation does not fluctuate significantly, it implies the existence of a very large ESR.

\begin{figure}
\centering
\includegraphics[trim = 0mm 0mm 0mm 0mm, clip, width=0.45\textwidth]{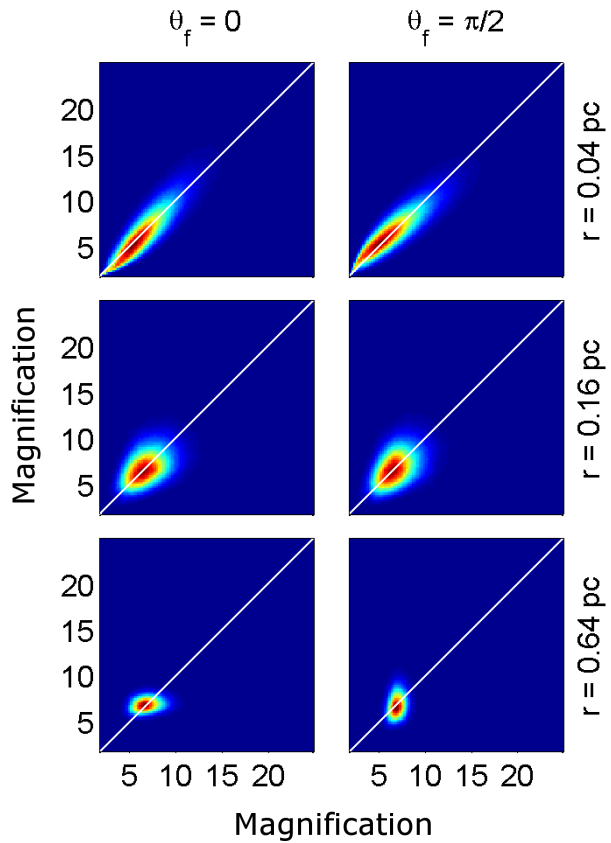}
\caption{Joint distributions for the magnifications of the VP filter (vertical axes) versus the magnifications of the HP profile (horizontal axes), for image D with unrotated maps. The rotation of the polarising filter and the ESR radius vary, with the ESR width set to $0.04$ pc. The columns correspond to relative orientations of the polarisation filter and the rows correspond to the inner radius of the ESR annulus. White lines denote equal magnification of both components.}
\label{Fig:Preferencing}
\end{figure}

With the spatial separation between HP and VP wings of the model ESR annulus, as evident in Figure \ref{Fig:ModelExample}, there is an expectation that the caustic structure of magnification maps would preference either the magnification of horizontally or vertically polarised light. The symmetry of distributions in Figs. \ref{Fig:RotationDistributionsB} and \ref{Fig:ABCDDistributionComparisons} suggests that no such effect is discernible for a default inner radius of $0.08$ pc and a width of $0.32$ pc. However, the low ESR-width panels of Figure \ref{Fig:SizeComparisons} demonstrate that there is a clear asymmetric preferencing effect if the ESR annulus is sufficiently thin.

Figure \ref{Fig:Preferencing} investigates this further by fixing an ESR width of $0.04$ pc and varying the inner radius. The left column has $\theta_f = 0$ so the HP wings are parallel to the caustic bands. The right column has $\theta_f = \pi/2$ so the HP wings are perpendicular to the bands. It is evident in the bottom two panels that sufficiently large inner radii allow greatest fluctuations in magnification for the parallel components. This is because the perpendicular wings are less influenced by perturbations from traversing individual caustic bands; they lie across a spread of caustics and only have a relatively small portion of their area crossing new bands as the lens moves. The significance of this is that the shape of the joint distributions gathered over a substantial period of time will potentially describe both the width of the ESR and its distance from the accretion disk.

\subsection{Time-Series Statistics}
\label{Subsection:TemporalStatistics}
In  Q2237+0305 the estimated transverse velocity of the lensing galaxy is $600$ km s$^{-1}$ \citep{Wambsganss:90a}; leading to a projected velocity across the source plane of $\sim 6500$ km s$^{-1}$ \citep[see][]{1986A&A...166...36K}. It is this large projected velocity, and correspondingly small variability time-scale, that makes Q2237+0305 such an ideal laboratory in which to study gravitational microlensing. Using this, time-series statistics that predict periodicities can be tested and the structure of a quasar's ESR may be resolved by comparing simulated results with observations.

\begin{figure}
\centering
\includegraphics[trim = 0mm 0mm 0mm 0mm, clip, width=0.5\textwidth]{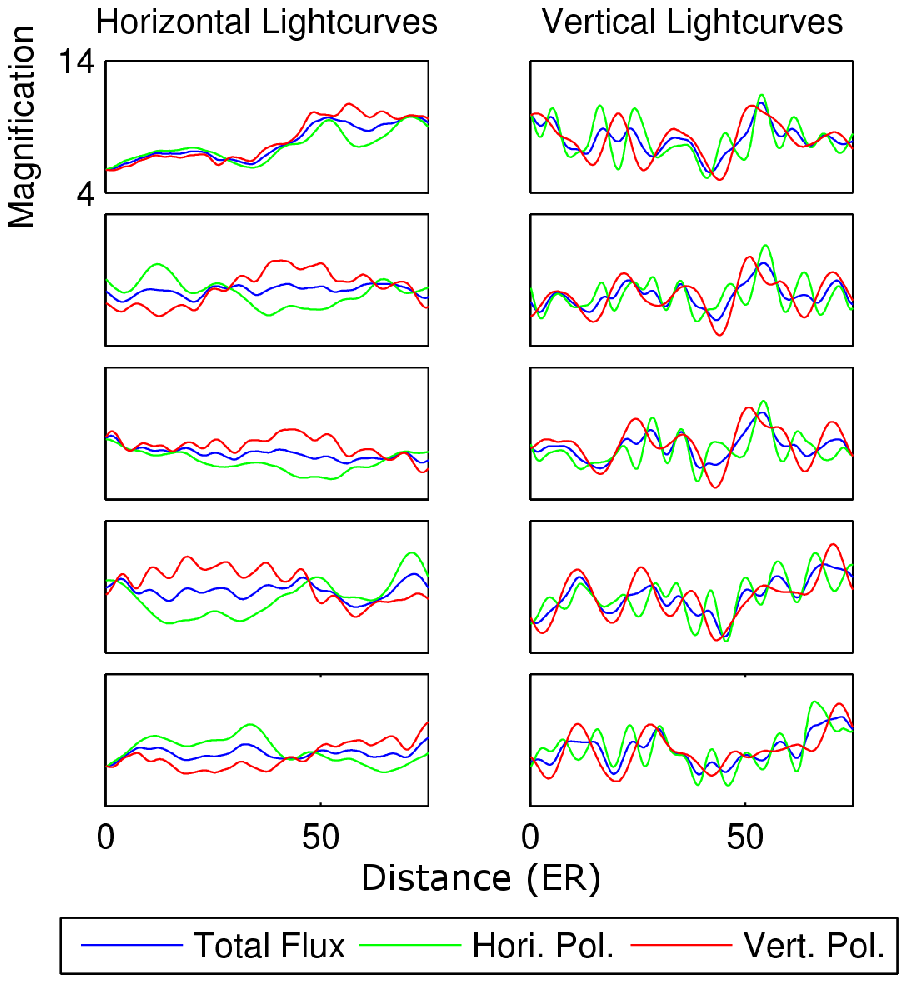}
\caption{Light curves for the base ESR and image A, using a single unrotated map and unrotated polarisation filter. The left panel shows five light curves from horizontal paths, the right shows five vertical paths. The paths were separated by $1/4$ of the map width. The blue line represents the source profile for the total annulus flux, the green line
for the HP filter, and the red line for the VP filter. Distance units are in Einstein Radii.}
\label{Fig:LightCurve}
\end{figure}

The primary tool for temporal statistics in microlensing is the light curve, as seen in Figure \ref{Fig:LightCurve}. 
The left column displays five horizontal light curves for image A, $\phi = \theta_f = 0$, from a single map. The blue line represents the source profile for the total annulus flux, the green line relates to the HP filtered intensity, and the red line represents the VP filtered intensity. The right column is the same, except for vertical paths. By taking a cut through a generated convolution map, we represent the observed intensity of a source as its origin moves along the relevant path of the magnification map. It is reasonable to expect from the results gathered in Section \ref{Subsection:StaticStatistics} and Figure \ref{Fig:LightCurve} that fluctuations due to caustic crossings may be periodic in nature if the caustic networks themselves are spatially correlated.

\begin{figure}
\centering
\includegraphics[trim = 0mm 0mm 0mm 0mm, clip, width=0.45\textwidth]{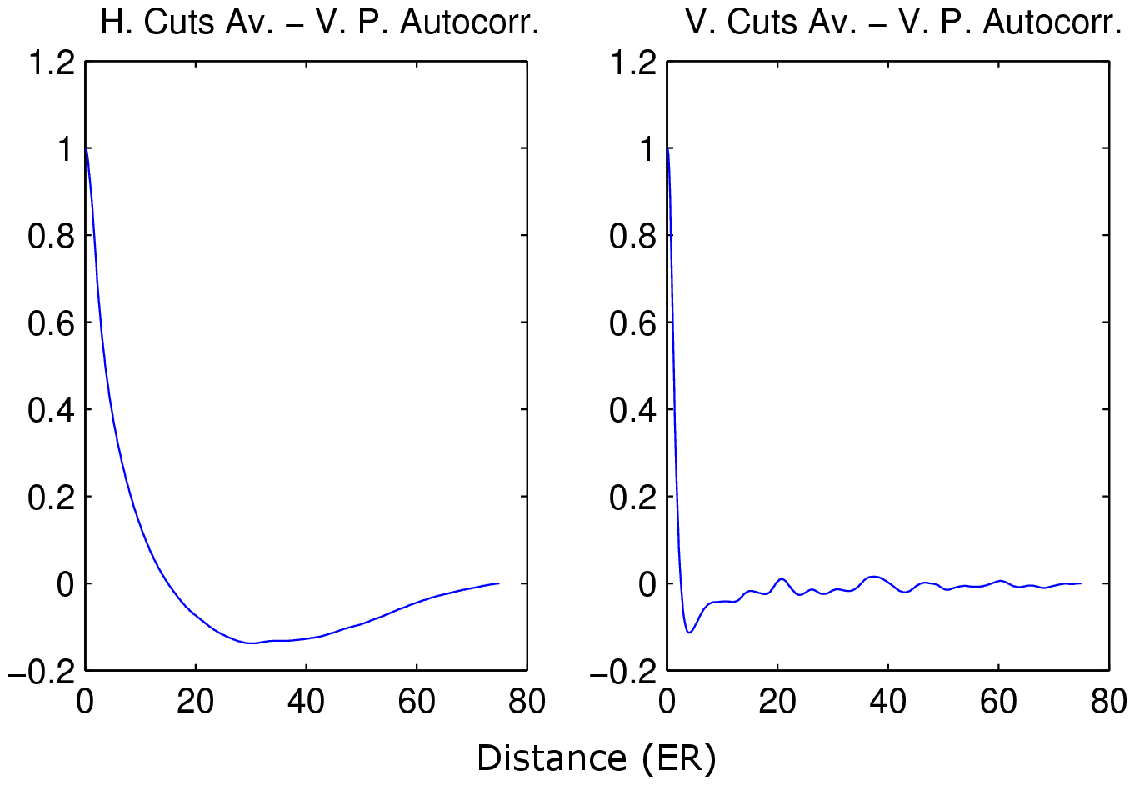}
\caption{Auto-correlation curves using the VP filter, with unrotated maps, for image B. A small ESR of $0.02$ pc for inner radius and $0.04$ pc for width is used as the source. All horizontal or vertical paths are used, and the results averaged. The left panel shows the results for horizontal light curves and the right panel displays the vertical analogue. The horizontal axis is in Einstein Radii and represents the lag used for auto-correlation measures. The vertical axis is in arbitrary units.}
\label{Fig:BaselineCorrelationCurve}
\end{figure}

We employ simple cross- and auto-correlation measures to examine whether periodicities occur in the light curves, which would indicate an approximately regular interaction between the caustic structure and both the size and orientation of the ESR \citep[see][for an in-depth discussion of microlensing correlation functions]{Seitz:94a,Seitz:94b}. Figure \ref{Fig:BaselineCorrelationCurve} shows auto-correlations of light curves for the VP filter of image B only, using horizontal map paths (left panel) and vertical map paths (right panel), and with $\phi = \theta_f = 0$. The source here is a small ESR with inner radius = $0.02$ pc and width = $0.04$ pc. Horizontal light curve paths are parallel to the caustic bands and this is reflected in the left panel which shows a smooth auto-correlation. Vertical paths are perpendicular to the caustic bands and show a less smooth auto-correlation. As the effect of relative rotation between shear axis and polarising filter merely reduces the degree of correlation, we will only display correlations for vertical light curves in what follows. Upon examination of the relevant correlation functions for the other images A, C and D (not shown), there appears to be repetitive structure within the light curves in steps of approximately $5.83$ ER ($0.35$ pc). Statistical analysis over more samples will be required to confirm this result.

\begin{figure*}
\centering
\includegraphics[trim = 0mm 0mm 0mm 0mm, clip, width=1\textwidth]{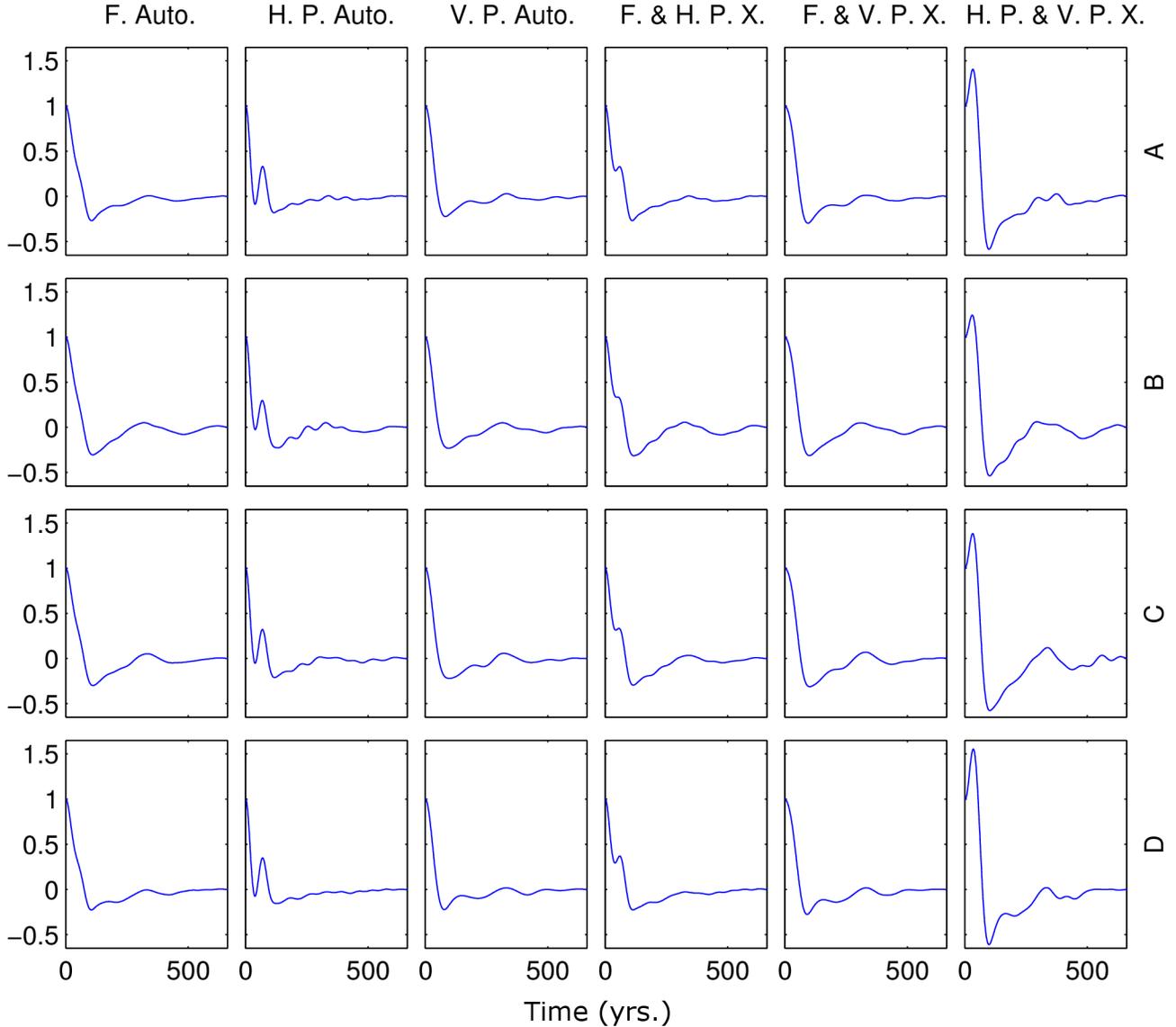}
\caption{Correlation curves for vertical light curve cuts for the base ESR, with unrotated maps and polarisation filter, for all images. From left to right, the columns display the magnification auto-correlation of total flux (annulus), HP filter, VP filter, and the cross-correlation between total flux and HP filter, total flux and VP filter, and between HP and VP filters. All possible vertical light curves in all unrotated maps were used, and the results averaged. ``F'' is total ESR flux (annulus), ``HP'' pertains to the HP filter, ``VP'' pertains to the VP filter, ``Auto'' indicates auto-correlation, and ``X'' indicates cross-correlation. The vertical axis is in arbitrary units and the horizontal axis is in years; for auto-correlation this represents the lag between light curves.}
\label{Fig:RealisticCorrelation}
\end{figure*}

Figure \ref{Fig:RealisticCorrelation} shows auto- and cross-correlations for vertical light curves from all images of the base ESR, using total flux and both the HP and VP filters, with $\phi = \theta_f = 0$. The rows represent different images and the columns different correlation measures. The meanings of the symbols are: ``F'' as total ESR flux (annulus), ``HP'' as the HP filter,
``VP'' as the VP filter, ``Auto'' as auto-correlation, and ``X'' as cross-correlation. These panels reveal that each of the four lens images generate similar correlations for the base ESR, indicating that convergence and shear values in this range do not affect the light curves in any significant way. Thus the regions of correlation are likely to be directly related to the shape of the model annulus. More importantly, the size of the source is estimable from the correlation curves, with the auto-correlation showing a distinct peak after a lag of approximately $71$ years ($0.48$ pc). This value is also the distance between the edge of the annulus that first encounters a caustic and the inner edge of the back half that follows, suggesting that variations in  magnification due to one point will be observed again when the source's back edge crosses the point.

The cross-correlation curves of Figure \ref{Fig:RealisticCorrelation} also give unusual peaks at approximately 327 years ($2.2$ pc), which may imply periodic structure in the caustic networks as the distance scale is larger than that pertaining to the source. More obvious though is the sharp peak correlating horizontally and vertically polarised magnification at distances of about $33$ years ($0.22$ pc), corresponding roughly to the vertical distance between the average position of the VP wings and either tip of the HP pair.

\begin{figure}
\centering
\subfloat[]{\label{Fig:SizeCorrelation}\includegraphics[trim = 0mm 0mm 0mm 0mm, clip, width=0.45\textwidth]{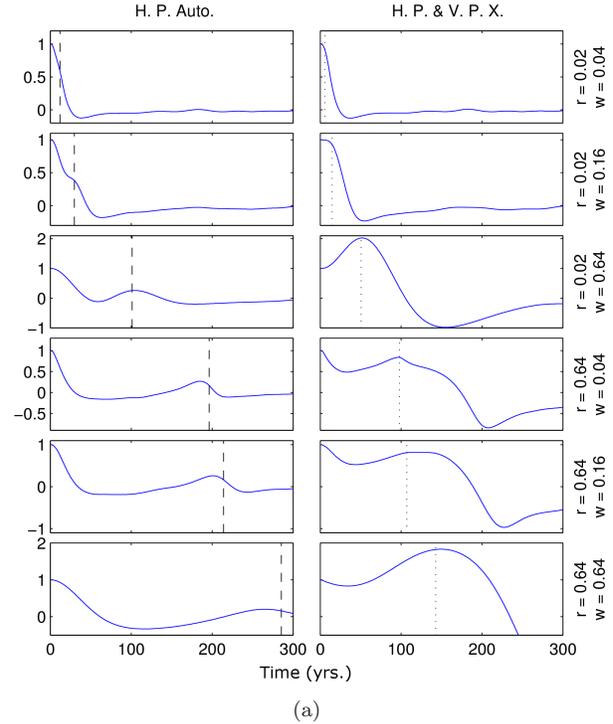}}\\
\subfloat[]{\label{Fig:CorrExplanation}\includegraphics[trim = 0mm 0mm 0mm 0mm, clip, width=0.45\textwidth]{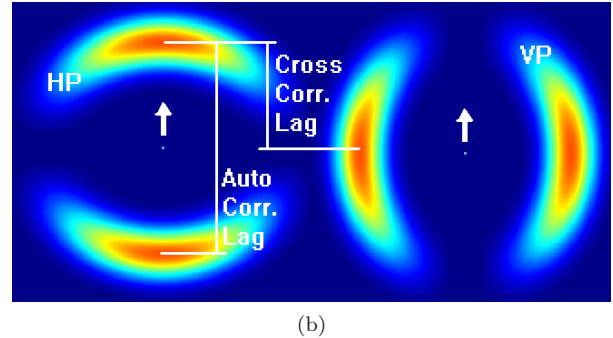}}
\caption{Correlation measures using vertical light curves, with unrotated maps and polarisation filter, for image D. The ESR model varies. In (a) the left panel is an auto-correlation using the HP filter, the right panel is a cross-correlation between HP and VP filters. ``Auto'' indicates auto-correlation, and ``X'' indicates cross-correlation. Dashed lines mark lags corresponding to the ESR diameter and dotted lines denote lags corresponding to ESR radius ($r_E+w_E/2$ according to Section~\ref{Subsection:Model}). The vertical axis is in arbitrary units and horizontal axis is in years. (b) displays the ESR distances that relate to the lags for the cross- and auto-correlation in (a).}
\label{Fig:SizeCorrelations}
\end{figure}

To further investigate whether the correlation features corresponding to ESR size are general, Figure \ref{Fig:SizeCorrelation} shows correlation measures and Figure \ref{Fig:CorrExplanation} displays a diagram indicating the ESR separations that relate to the lags for the cross- and auto-correlations in Figure \ref{Fig:SizeCorrelation}. The left panel in Figure \ref{Fig:SizeCorrelation} shows the HP filter light curve auto-correlation and the right panel shows the cross-correlation of HP filter light curves with VP filter light curves; all with $\phi = \theta_f = 0$. ``Auto'' indicates auto-correlation, and ``X'' indicates cross-correlation. Overlaying each plot is either a dashed line that corresponds to the time the ESR takes to travel its diameter ($2r_E+w_E$) or a dotted line that relates to the radius ($r_E+w_E/2$), both defined graphically in Figure \ref{Fig:CorrExplanation}. There is good agreement between both the lines and the correlation peaks; if one HP wing crosses a caustic, it is highly probable the second will too once the ESR travels the length of its diameter. Likewise, the perpendicular VP wings will also cross the caustic once the ESR travels the length of its diameter.

The curvature of the wings at larger inner radii and widths means that the correlation length is  lower than the ESR diameter, shown by the weakened correlation between dashed lines and peaks in the bottom three panels of Figure \ref{Fig:SizeCorrelation}. More detailed calculations relating to the spatial mean of wing intensity will improve the agreement. However, it is evident that the size of an ESR according to our model is discernible from the relevant correlation curves.

\section{Conclusion}
\label{Section:Conclusion}
In this paper we have presented a numerical study of the influence of gravitational microlensing upon the electron scattering region thought to exist in the inner regions of quasars. Focusing upon the quadruply imaged quasar, Q2237+0305, we considered a fiducial model with a $0.08$ pc inner radius and a $0.32$ pc width consistent with recent observations \citep{Taniguchi:99}. As seen in Figure \ref{Fig:RotationDistributionsB}, the magnification of the two polarised annulus components can also be well correlated if the relative angle between filter and caustics is an odd multiple of $\pi/4$. It was also discovered that, by orienting a polarising filter and defining an axis of polarisation as parallel to the magnification map's shear axis, larger ESR dimensions allow for an increased level of events where one polarisation is magnified more than the other.

By considering the cross- and auto-correlation of the resulting light curves, it is seen that periodicities in the microlensing magnification maps are imprinted on the light curve, providing further clues to the ESR scattering geometry, with the size of the ESR being discernible from the caustic-crossing auto-correlation curve of the polarised ``wings'' that are parallel to the caustics, or the cross-correlation between the magnification of both polarisations.

Polarimetric observations during a known microlensing event would provide more information than constant polarimetric monitoring, which is observationally more expensive. Q2237+0305 is regularly monitored photometrically \citep[e.g. the OGLE project]{Udalski:2006} and a photometric microlensing event could ``trigger'' an intense polarimetric study [such overrides have been proposed for Q2237+0305 in the past \citep[e.g.][]{Webster:2004}], which would directly probe the structure of the ESR.

It must be noted that, to obtain long term variability statistics, photometric and subsequent polarimetric monitoring will be required on decade to century time-scales, even to observe the characteristic correlation lags indicated in Figure \ref{Fig:SizeCorrelation} for the small-scale fiducial model. Nevertheless, a monitoring campaign on that order of duration would already be able to compile a magnification range indicative of ESR size, as suggested by Figure \ref{Fig:SizeComparisons}, especially for a small inner radius. Constraining a large-scale ESR should be possible too, as the preferencing effects for polarizing filter rotations, as shown in Figure \ref{Fig:Preferencing}, are likely to be observed over the same order of time.

For this study, we considered a fiducial physically motivated model of the ESR, represented as a circular torus about the central accretion disk. While we have examined some of the freedom of this model, there is substantial scope to extend considerations to more complex geometries (such as a tilting of the emission regions relative to the line of sight). We will consider these in further contributions.

\section*{Acknowledgments}
We thank the anonymous reviewer whose comments improved the content and quality of the paper.
Computing facilities were provided by the High Performance Computing Facility, University of Sydney. This work is undertaken as part of the Commonwealth Cosmology Initiative (www.thecci.org), and funded by the Australian Research Council Discovery Project DP0665574.

\bsp

\label{lastpage}


\begin{thebibliography}{99}

\bibitem[\protect\citeauthoryear{Abajas et al.}{2002}]{Abajas:2002} 
Abajas C., Mediavilla E., Mu{\~n}oz J.~A., Popovi{\'c} L.~{\v C}., Oscoz 
A., 2002, ApJ, 576, 640 

\bibitem[\protect\citeauthoryear{Abajas et al.}{2007}]{Abajas:2007} 
Abajas C., Mediavilla E., Mu{\~n}oz J.~A., G{\'o}mez-{\'A}lvarez P., 
Gil-Merino R., 2007, ApJ, 658, 748 

\bibitem[\protect\citeauthoryear{Agol et al.}{2009}]{Agol:09} 
Agol E., Gogarten S.~M., Gorjian V., Kimball A., 2009, ApJ, 697, 1010 

\bibitem[\protect\citeauthoryear{Antonucci}{1999}]{Antonucci:99} 
Antonucci R., 1999, ASPC, 161, 193 

\bibitem[\protect\citeauthoryear{Bate et al.}{2008}]{Bate:2008} 
Bate N.~F., Floyd D.~J.~E., Webster R.~L., Wyithe J.~S.~B., 2008, MNRAS, 
391, 1955 

\bibitem[\protect\citeauthoryear{Belle \& Lewis}{2000}]{2000PASP..112..320B} Belle K.~E., Lewis G.~F., 2000, PASP, 112, 320 

\bibitem[\protect\citeauthoryear{Blackburne et al.}{2010}]{Pooley:2010} Blackburne J.~A., Pooley D., Rappaport S., 
Schechter P.~L., 2010, arXiv, arXiv:1007.1665 

\bibitem[\protect\citeauthoryear{Brewer \& Lewis}{2005}]{Brewer:2005} Brewer B.~J., Lewis G.~F., 2005, MNRAS, 356, 703 

\bibitem[\protect\citeauthoryear{Chae et al.}{2001}]{Chae:2001} 
Chae K.-H., Turnshek D.~A., Schulte-Ladbeck R.~E., Rao S.~M., Lupie O.~L., 2001, ApJ, 561, 653 

\bibitem[\protect\citeauthoryear{Draine}{2003}]{Draine:03} Draine 
B.~T., 2003, ApJ, 598, 1026 

\bibitem[\protect\citeauthoryear{Elvis}{2000}]{Elvis:00} Elvis 
M., 2000, NewAR, 44, 559 

\bibitem[\protect\citeauthoryear{Garsden \& Lewis}{2010}]{2010NewA...15..181G} Garsden H., Lewis G.~F., 2010, NewA, 15, 181 

\bibitem[\protect\citeauthoryear{Hales \& Lewis}{2007}]{Hales:07} Hales C.~A., Lewis G.~F., 2007, PASA, 24, 30

\bibitem[\protect\citeauthoryear{Huchra et al.}{1985}]{Huchra:85} 
Huchra J., Gorenstein M., Kent S., Shapiro I., Smith G., Horine E., Perley 
R., 1985, AJ, 90, 691 

\bibitem[\protect\citeauthoryear{Kayser, Refsdal, \& Stabell}{1986}]{1986A&A...166...36K} Kayser R., Refsdal S., Stabell R., 1986, A\&A, 166, 36 

\bibitem[\protect\citeauthoryear{Kayser \& Refsdal}{1989}]{Kayser:89} Kayser R., Refsdal S., 1989, Natur, 338, 745 

\bibitem[\protect\citeauthoryear{Keeton et al.}{2006}]{Keeton:2006} 
Keeton C.~R., Burles S., Schechter P.~L., Wambsganss J., 2006, ApJ, 639, 1 

\bibitem[\protect\citeauthoryear{Kishimoto et 
al.}{2008a}]{Kishimoto:08a} Kishimoto M., Antonucci R., Blaes O., 
Lawrence A., Boisson C., Albrecht M., Leipski C., 2008, JPhCS, 131, 012039 

\bibitem[\protect\citeauthoryear{Kishimoto et 
al.}{2008b}]{Kishimoto:08b} Kishimoto M., Antonucci R., Blaes O., 
Lawrence A., Boisson C., Albrecht M., Leipski C., 2008, Nature, 454, 492

\bibitem[\protect\citeauthoryear{Lewis \& Irwin}{1995}]{Lewis:95a} Lewis G.~F., Irwin M.~J., 1995, MNRAS, 276, 103 

\bibitem[\protect\citeauthoryear{Lewis \& Ibata}{2004}]{Lewis:2004} Lewis G.~F., Ibata R.~A., 2004, MNRAS, 348, 24 

\bibitem[\protect\citeauthoryear{Nemiroff}{1988}]{1988ApJ...335..593N} 
Nemiroff R.~J., 1988, ApJ, 335, 593 

\bibitem[\protect\citeauthoryear{Peterson}{1998}]{1998AdSpR..21...57P} 
Peterson B.~M., 1998, AdSpR, 21, 57

\bibitem[\protect\citeauthoryear{Richards et al.}{2004}]{Richards:2004} Richards G.~T., et al., 2004, ApJ, 610, 
679 

\bibitem[\protect\citeauthoryear{Refsdal \& Stabell}{1991}]{Refsdal:91} Refsdal S., Stabell R., 1991, A\&A, 250, 62 

\bibitem[\protect\citeauthoryear{Rhie \& Bennett}{1999}]{Rhie:99} Rhie S.~H., Bennett D.~P., 1999, astro, arXiv:astro-ph/9912050 

\bibitem[\protect\citeauthoryear{Rix, Schneider, \& Bahcall}{1992}]{1992AJ....104..959R} Rix H.-W., Schneider D.~P., Bahcall J.~N., 1992, AJ, 104, 959 

\bibitem[\protect\citeauthoryear{Schild \& Vakulik}{2003}]{Schild:03} Schild R., Vakulik V., 2003, AJ, 126, 689 

\bibitem[\protect\citeauthoryear{Schneider, Ehlers, \& Falco}{1992}]{Schneider:92}
Schneider P, Ehlers J, Falco EE. Gravitational Lenses. Springer-Verlag: Berlin 1992.

\bibitem[\protect\citeauthoryear{Schneider, Kochanek, \& Wambsganss}{2006}]{2006glsw.book.....S} Schneider P., Kochanek C.~S., Wambsganss J., 2006,  
Saas-Fee Advanced Courses, Springer-Verlag Berlin Heidelberg

\bibitem[\protect\citeauthoryear{Schneider \& Wambsganss}{1990}]{1990A&A...237...42S} Schneider P., Wambsganss J., 1990, A\&A, 237, 42 

\bibitem[\protect\citeauthoryear{Seitz \& Schneider}{1994}]{Seitz:94a} Seitz C., Schneider P., 1994, A\&A, 288, 1 

\bibitem[\protect\citeauthoryear{Seitz, Wambsganss, \& Schneider}{1994}]{Seitz:94b} Seitz C., Wambsganss J., Schneider P., 1994, A\&A, 288, 19 

\bibitem[\protect\citeauthoryear{Taniguchi \& Anabuki}{1999}]{Taniguchi:99} Taniguchi Y., Anabuki N., 1999, ApJ, 521, L103 

\bibitem[\protect\citeauthoryear{Udalski et al.}{2006}]{Udalski:2006} Udalski A., et al., 2006, AcA, 56, 293 

\bibitem[\protect\citeauthoryear{Wambsganss, Paczynski, \& Katz}{1990}]{Wambsganss:1990b} Wambsganss J., Paczynski B., Katz N., 1990, ApJ, 352, 407 

\bibitem[\protect\citeauthoryear{Wambsganss, Paczynski, \& Schneider}{1990}]{Wambsganss:90a} Wambsganss J., Paczynski B., Schneider P., 1990, ApJ, 358, L33 

\bibitem[\protect\citeauthoryear{Wambsganss}{1999}]{Wambsganss:90}
Wambsganss J., 1999, JCoAM, 109, 353 

\bibitem[\protect\citeauthoryear{Wayth, O'Dowd, \& Webster}{2005}]{Wayth:2005} Wayth R.~B., O'Dowd M., Webster R.~L., 2005, MNRAS, 359, 561 

\bibitem[\protect\citeauthoryear{Webster}{2004}]{Webster:2004} 
Webster R., 2004, hst..prop, 10123 

\bibitem[\protect\citeauthoryear{Witt}{1990}]{1990A&A...236..311W} Witt H.~J., 1990, A\&A, 236, 311 

\bibitem[\protect\citeauthoryear{Witt, Kayser, \& Refsdal}{1993}]{Witt:1993} Witt H.~J., Kayser R., Refsdal S., 1993, A\&A, 268, 501 

\bibitem[\protect\citeauthoryear{Wo{\'z}niak et al.}{2000}]{Wozniak:2000} Wo{\'z}niak P.~R., Alard C., Udalski A., 
Szyma{\'n}ski M., Kubiak M., Pietrzy{\'n}ski G., Zebru{\'n} K., 2000, ApJ, 
529, 88 

\bibitem[\protect\citeauthoryear{Wyithe}{2001}]{Wyithe:2001} Wyithe J.~S., 2001, ASPC, 237, 201 

\bibitem[\protect\citeauthoryear{Yonehara}{2001}]{Yonehara:2001} 
Yonehara A., 2001, ApJ, 548, L127 

\end{thebibliography}
\end{document}